\documentstyle[12pt]{article}                    
\newfont{\ffont}{msym10}                          
\newcommand{\beq}{\begin{equation}}               
\newcommand{\eeq}{\end{equation}}                 
\newcommand{\bqry}{\begin{eqnarray}}              
\newcommand{\eqry}{\end{eqnarray}}                
\newcommand{\bqryn}{\begin{eqnarray*}}            
\newcommand{\eqryn}{\end{eqnarray*}}              
\newcommand{\NL}{\nonumber \\}                    
\newcommand{\PD}[2]                               
    {\frac{\partial^{#2}}{\partial #1^{#2}}}      
\begin{document}
\title{Towards Resolution of the Scalar Meson Nonet Enigma 
II.Gell-Mann--Okubo Revisited}
\author{\\ L. Burakovsky\thanks{E-mail: BURAKOV@PION.LANL.GOV} \
and \ T. Goldman\thanks{E-mail: GOLDMAN@T5.LANL.GOV} \
\\  \\  Theoretical Division, MS B285 \\  Los Alamos National Laboratory \\ 
Los Alamos, NM 87545, USA \\}
\date{ }
\maketitle
\begin{abstract}
The new SU(3) nonet mass formula $2M^2(s\bar{s})+3M^2(n\bar{n},I=1)=4M^2(s
\bar{n})+M^2(n\bar{n},I=0)$ $(n=u,d),$ obtained in our previous paper by using
Regge phenomenology, is rederived for the pseudoscalar and scalar mesons in the
Nambu--Jona-Lasinio model with instanton-induced interaction and applied to the
problem of the correct $q\bar{q}$ assignment for the scalar meson nonet. The 
results strongly favor the masses of the scalar isoscalar mostly octet and 
mostly singlet states in the vicinity of 1.45 GeV and 1.1 GeV, respectively.   
\end{abstract}

\section*{ }
The spectrum of the scalar meson nonet is a long-standing problem of light 
meson spectroscopy. The number of resonances found in 
the region of 1--2 GeV exceeds the number of states that conventional quark 
models can accommodate \cite{pdg}. Extra states are interpreted alternatively 
as $K\bar{K}$ molecules, glueballs, multi-quark states or hybrids. In 
particular, except for a well established scalar isodoublet state, the $K_0^
\ast (1430),$ the Particle Data Group (PDG) \cite{pdg} lists two isovector 
states, the $a_0(980)$ and $a_0(1450).$ The latter, having mass and width 
1450$\pm 40$ MeV, $270\pm 40$ MeV, respectively, was discovered recently by 
the Crystal Barrel collaboration \cite{LEAR}. A third isovector state (not 
included in \cite{pdg}), $a_0(1320),$ having mass and width $1322\pm 30$ MeV 
and $130\pm 30$ MeV, was seen by GAMS \cite{GAMS} and LASS \cite{Aston} in the
partial wave analyses of the $\eta \pi $ and $K_sK_s$ data, respectively. 

There are four isoscalar states in \cite{pdg}, the $f_0(400-1200)$ (or $\sigma
),$ the interpretation of which as a particle is controversial due to a huge 
width of 600--1000 MeV, $f_0(980),$ $f_0(1370)$ (which stands for two separate
states, $f_0(1300)$ and $f_0(1370)$, of a previous edition of PDG \cite{pdg1}),
and $f_0(1500)$ (which also stands for two separate states, $f_0(1525)$ and 
$f_0(1590),$ of a previous edition of PDG), and two more possibly scalar 
states, the $f_J(1710),$ $J=0$ or 2, seen in radiative $J/\Psi $ decays, and 
an $\eta $-$\eta $ resonance $X(1740)$ with uncertain spin, produced in 
$p\bar{p}$ annihilation in flight and in charge-exchange. Recently several 
groups claimed different scalar isoscalar structures close to 1500 MeV, 
including: 1) a narrow state with mass $1445\pm 5$ MeV and width $65\pm 10$ 
MeV seen by the WA91 collaboration at CERN in central production of $4\pi $ in
high-energy $pp$ collisions \cite{Abatzis}, 2) the lightest of the three 
states with masses 1505 MeV, 1750 MeV and 2104 MeV revealed upon reanalyzing 
of data on $J/\Psi \rightarrow \gamma 2\pi ^{+}2\pi ^{-}$ \cite{Bugg}, and 3)
the $f_0(1400),$ $f_0(1500),$ $f_0(1520).$ The masses, widths and decay 
branching ratios of these states are incompatible within the errors quoted by 
the groups. We do not consider it as plausible that so many scalar isoscalar 
states exist in such a narrow mass interval. Instead, we take the various 
states as manifestation of one object which we identify tentatively with the 
$f_0(1450).$

It has been convincingly argued that the narrow $a_0(980)$, which has also 
been seen as a narrow structure in $\eta \pi $ scattering, can be generated by
meson-meson dynamics alone \cite{WI,KK}. This interpretation of the $a_0(980)$ 
leaves the $a_0(1320)$ or $a_0(1450)$ (which may be manifestations of one 
state having a mass in the interval 1350-1400 MeV) as the 1 $^3P_0$ $q\bar{q}$
state. Similarly, it is usually assumed that the $f_0(980)$ is a $K\bar{K}$ 
molecule, as suggested originally by Weinstein and Isgur \cite{WI}. The mass 
degeneracy with the $a_0(980)$ and their proximity to the $K\bar{K}$ threshold
seem to require that the nature of both states should be the same. On the 
other hand, the $K\bar{K}$ interaction in the $I=1$ and $I=0$ channels is very
different: the extremely attractive $I=0$ interaction may not support a 
loosely bound state. Instead, it may just define the pole position of the 
$f_0(980)$ $q\bar{q}$ resonance. Indeed, Morgan and Pennington \cite{MP} find 
the $f_0(980)$ pole structure characteristic for a genuine resonance of the 
constituents and not of a weakly bound system. The $I=1$ $K\bar{K}$ interaction
is weak and may generate a $K\bar{K}$ molecule. Alternatively, T\"{o}rnqvist 
\cite{Torn} interprets both the $f_0(980)$ and $a_0(980)$ as the members of the
$q\bar{q}$ nonet with strong coupling to the decay channels. However, this
does not account for the recently discovered $a_0(1320)$ and $a_0(1450).$

With respect to the $f_0(1370)$ (or two separate states, $f_0(1300)$ and $f_0(
1370),$ according to a previous edition of PDG), one may follow the arguments 
of Morgan and Pennington \cite{MP} and assume that the $\pi \pi $ interaction 
produces both very broad, $f_0(1000),$ and narrow, $f_0(980),$ states, giving 
rise to a dip at 980 MeV in the squared $\pi \pi $ scattering amplitude $T_{11
}.$ In this picture, the $f_0(1370)$ is interpreted as the high-mass part of 
the $f_0(1000)$ (the low-mass part may be associated with the $\sigma $ of the
most recent PDG). In experiments, the $f_0(1000)$ shows up at $\sim 1300$ MeV 
because of the pronounced dip in $|T_{11}|^2$ at $\sim 1$ GeV. The $f_0(1000)$
has an extremely large width; thus a resonance interpretation is 
questionable. It could be generated by $t$-channel exchanges instead of 
inter-quark forces \cite{ZB}.

The $f_0(1500)$ resonance has been recently observed by the Crystal Barrel
collaboration in $p\bar{p}$ annihilations \cite{CrBa}. It was claimed that 
this state has a peculiar decay pattern\footnote{New preliminary results by
Crystal Barrel are \cite{Koch} $$\pi \pi :\eta \eta :\eta \eta ^{'}:K\bar{K}=
1.45:0.34:0.48:0.48\pm 0.24,$$ with normalization of $\pi \pi $ in agreement 
with (1).} \cite{AC} 
\beq
\pi \pi :\eta \eta :\eta \eta ^{'}:K\bar{K}=1.45:0.39\pm 0.15:0.28\pm 0.12:
<0.15. 
\eeq
This pattern can be reproduced by assuming the existence of an additional 
scalar state which is mainly $s\bar{s}$ and should have a mass of about 1700 
MeV, possibly the $f_J(1710),$ and tuning the mixing of the $f_0(1500)$ with 
the $f_0(1370)$ $n\bar{n}$ $(n=u,d)$ and the (predicted) $f_0(1700)$ $s\bar{s}$
states \cite{AC}. In this picture, the $f_0(1500)$ is interpreted as a glueball
state with strong mixing with the close-by conventional scalar mesons. 

On the other hand, Lee and Weingarten \cite{LW} interpret the $f_0(1500)$ as a
mainly $s\bar{s}$ state which mixes strongly with the close-by mainly $n\bar{
n}$ and scalar glueball states which show up as resonances at 1390 MeV and 1710
MeV, respectively. The latter is in agreement with the values for the scalar 
glueball mass $1740\pm 71$ MeV and $1710\pm 50$ MeV obtained from QCD lattice 
calculations by Sexton {\it et al.} \cite{SVW} and Luo {\it et al.} \cite{Luo},
respectively. An interpretation of the $f_0(1500)$ as a conventional $q\bar{q}$
state, as well as a qualitative explanation of its reduced $K\bar{K}$ partial 
width, were also given by Klempt {\it et al.} \cite{Klempt} in a relativistic 
quark model with linear confinement and instanton-induced interaction. A 
quantitative explanation of the reduced $K\bar{K}$ partial width of the $f_0(
1500)$ was given in a recent publication by the same authors \cite{Klempt1}.
The decay pattern obtained in \cite{Klempt1} (with the $f_0$-$f_0^{'}$ mixing 
angle $\approx 25^o),$ $$\pi \pi :\eta \eta :\eta \eta ^{'}:K\bar{K}=1.45:0.32
:0.18:0.03,$$ is in excellent agreement with (1).  

The above arguments lead one to the following spectrum of the scalar meson 
nonet (in the order: isovector, isodoublet, isoscalar mostly octet,
isoscalar mostly singlet),
\beq
a_0(1320)\;\;{\rm or}\;\;a_0(1450),\;\;\;K_0^\ast (1430),\;\;\;f_0(1500),\;\;\;
f_0(980)\;\;{\rm or}\;\;f_0(1000).
\eeq
This spectrum agrees essentially with the $q\bar{q}$ assignments found by 
Klempt {\it et al.} \cite{Klempt}, and Dmitrasinovic \cite{Dmitra} who
considered the Nambu--Jona-Lasinio model with a $U_A(1)$ breaking 
instanton-induced 't Hooft interaction. The spectrum of the 
meson nonet given in \cite{Klempt} is
\beq
a_0(1320),\;K_0^\ast (1430),\;f_0(1470),\;f_0(980),
\eeq
while that suggested by Dmitrasinovic, on the basis of the sum rule
\beq
m_{f_0}^2+m_{f_0^{'}}^2+m_\eta ^2+m_{\eta ^{'}}^2=2(m_K^2+m_{K_0^\ast }^2)
\eeq
derived in his paper, is \cite{Dmitra}
\beq
a_0(1320),\;K_0^\ast (1430),\;f_0(1590),\;f_0(1000).
\eeq
The $q\bar{q}$ assignment obtained by one of the authors by the application of
the linear mass spectrum discussed in ref. \cite{linear} to a composite system
of the two, pseudoscalar and scalar nonets, is \cite{invited}
\beq
a_0(1320),\;K_0^\ast (1430),\;f_0(1525),\;f_0(980),
\eeq
in essential agreement with (3) and (5). The assignment (6) has found further 
justification in the constituent quark model explored by us in ref. 
\cite{P-wave}.  

In our previous paper \cite{BG}, by using Regge phenomenology, we derived a 
new mass relation $(I$ stands for isospin):
\beq
2M^2(s\bar{s})+3M^2(n\bar{n},I=1)=4M^2(s\bar{n})+M^2(n\bar{n},I=0).
\eeq
This relation was obtained for the pseudoscalar mesons, and further generalized
to every meson multiplet. It differs from the Sakurai mass formula 
\cite{Sakurai} obtained in the case of the ideal nonet mixing,
\beq
\theta _{id}=\arctan \frac{1}{\sqrt{2}}\approx 35.3^o,
\eeq
\beq
2M^2(s\bar{s})+M^2(n\bar{n},I=1)+M^2(n\bar{n},I=0)=4M^2(s\bar{n})
\eeq
by only a term which depends explicitly on isospin variation, $\sim (M^2(n\bar{
n},I=1)-M^2(n\bar{n},I=0)).$ This term improves the accuracy of the Sakurai 
formula, which is not bad by itself, $(\sim $2-3\%) by a factor of 2 \cite{BG}.

In this paper we present another derivation of the formula (7) for 
pseudoscalar and scalar mesons in the Nambu--Jona-Lasinio model with an
instanton-induced interaction. Before doing this, let us mention that this 
formula can be applied to the physical states of these two nonets provided the 
mixing angles are known. For example, for pseudoscalar mesons, the 
$\eta $-$\eta ^{'}$ mixing angle is given by duality constraints \cite{Bra},
\beq
\tan \theta _{\eta \eta ^{'}}=-\frac{1}{2\sqrt{2}},\;\;\;\theta _{\eta \eta ^{
'}}\approx -19.5^o,
\eeq
in good agreement with most of experimental data \cite{data}. In view of the 
relations 
$$\left(
\begin{array}{c}
\eta  \\
\eta ^{'}
\end{array}
\right) =\left(
\begin{array}{lr}
\cos \theta _{\eta \eta ^{'}} & -\sin \theta _{\eta \eta ^{'}} \\
\sin \theta _{\eta \eta ^{'}} & \cos \theta _{\eta \eta ^{'}}
\end{array}
\right) \left(
\begin{array}{c}
\eta _8 \\
\eta _9
\end{array}
\right) ,\;\;\left(
\begin{array}{c}
\eta _s \\
\eta _n
\end{array}
\right) =\left(
\begin{array}{lr}
\cos \theta _{id} & -\sin \theta _{id} \\
\sin \theta _{id} & \cos \theta _{id}
\end{array}
\right) \left(
\begin{array}{c}
\eta _8 \\
\eta _9
\end{array}
\right) ,$$ where 
$$\eta _8=\frac{u\bar{u}+d\bar{d}-2s\bar{s}}{\sqrt{6}},\;\;\;\eta _9=\frac{
u\bar{u}+d\bar{d}+s\bar{s}}{\sqrt{3}}$$ are the isoscalar octet and singlet
states, respectively, and
$$\eta _n=\frac{u\bar{u}+d\bar{d}}{\sqrt{2}},\;\;\;\eta _s=s\bar{s}$$ are the
``ideal-mixture'' counterparts of the physical $\eta $ and $\eta ^{'}$ states,
one obtains
$$\left(
\begin{array}{c}
\eta _s \\
\eta _n
\end{array}
\right) =\left(
\begin{array}{lr}
\cos (\theta _{id}-\theta _{\eta \eta ^{'}}) & -\sin (\theta _{id}-\theta _{
\eta \eta ^{'}}) \\
\sin (\theta _{id}-\theta _{\eta \eta ^{'}}) & \cos (\theta _{id}-\theta _{
\eta \eta ^{'}})
\end{array}
\right) \left(
\begin{array}{c}
\eta  \\
\eta ^{'}
\end{array}
\right) $$
\beq
=\left(
\begin{array}{lr}
\cos \xi  & -\sin \xi \\
\sin \xi  & \cos \xi
\end{array}
\right) \left(
\begin{array}{c}
\eta  \\
\eta ^{'}
\end{array}
\right) ,
\eeq
where $\xi \equiv \theta _{id}-\theta _{\eta \eta ^{'}}$ and, as follows from 
(8), 
\beq
\cos \xi =\frac{\sin \theta _{\eta \eta ^{'}}+\sqrt{2}\cos \theta _{\eta \eta 
^{'}}}{\sqrt{3}}.
\eeq

Assuming, as usual, that the relevant matrix elements are equal to the squared
masses of the corresponding states, and using the orthogonality of the $\eta $
and $\eta ^{'}$ as physical states, we obtain, using (10)-(12), 
\bqry
m^2_{\eta _n} & = & \frac{2}{3}\;m^2_{\eta }\;+\;\frac{1}{3}\;m^2_{\eta ^{'}},
 \\
m^2_{\eta _s} & = & \frac{1}{3}\;m^2_{\eta }\;+\;\frac{2}{3}\;m^2_{\eta ^{'}},
\eqry
in agreement with naive expectations from the quark content of these states
which is, in view of (10) \cite{BG},
$$\eta =\frac{u\bar{u}+d\bar{d}-s\bar{s}}{\sqrt{3}},\;\;\;\eta ^{'}=\frac{
u\bar{u}+d\bar{d}+2s\bar{s}}{\sqrt{6}}.$$
The use of the values (13),(14) in Eq. (7) leads finally to 
\beq
4m_K^2=3m_\pi ^2+m_{\eta ^{'}}^2,
\eeq
which is the new Gell-Mann--Okubo mass formula for pseudoscalar mesons found in
our previous publication \cite{BG} and which is satisfied to an accuracy of 
better than 1\% by the measured pseudoscalar meson masses.

We now turn to the derivation of the formula (7). We shall adopt the version 
of the Nambu--Jona-Lasinio (NJL) model which includes the U$_A(1)$ breaking 't
Hooft interaction \cite{tH} in the form of a $2N_f$-point determinant nonlocal
quark interaction, where $N_f$ is the number of flavors. This model has been
extensively studied by Dmitrasinovic \cite{Dmitra,Dmitra2}. For $N_f=2,$
and in the local interaction limit, the sum of the determinant and its 
Hermitian conjugate is equivalent to the following four-point interaction:
\bqry
L_{tH}^{(4)} & = & G_2\Big[ {\rm det}\left( \bar{\psi }(1+\gamma _5)\psi 
\right) +{\rm H.c.}\Big] \NL  
 & = & \frac{G_2}{2}\Big[ \left( \bar{\psi }\psi \right) ^2-\left( \bar{\psi }
\mbox{\boldmath $\tau $}\psi \right) ^2-\left( \bar{\psi }i\gamma _5\psi 
\right) ^2+\left( \bar{\psi }i\gamma _5\mbox{\boldmath $ \tau $}\psi \right) 
^2\Big] .
\eqry
Therefore, the resulting effective Lagrangian is
\bqry
L_{NJL}^{(4)} & = & L_{NJL}\;+\;L_{tH}^{(4)}\;=\;\bar{\psi }\left[ i\gamma 
\partial -m^0\right] \psi   \NL
 &   & +\;\frac{G_1}{2}\Big[ \left( \bar{\psi }\psi \right) ^2+\left( \bar{
\psi }\mbox{\boldmath $\tau $}\psi \right) ^2+\left( \bar{\psi }i\gamma _5\psi
\right) ^2+\left( \bar{\psi }i\gamma _5\mbox{\boldmath $\tau $}\psi \right) ^2
\Big]   \NL
 &   & +\;\frac{G_2}{2}\Big[ \left( \bar{\psi }\psi \right) ^2-\left( \bar{
\psi }\mbox{\boldmath $\tau $}\psi \right) ^2-\left( \bar{\psi }i\gamma _5\psi
\right) ^2+\left( \bar{\psi }i\gamma _5\mbox{\boldmath $\tau $}\psi \right) ^2
\Big] .
\eqry
The coupling constants $G_1$ and $G_2$ scale as $$G_1=O\left( \frac{1}{N_c}
\right) ,\;\;\;G_2=O\left( \frac{1}{N_c^2}\right) $$ in the large $N_c$ limit,
$N_c$ being the number of colors \cite{Dmitra}. Note that in the point limit 
of single gluon exchange, QCD would produce the $G_1$ temrs only (as well as
their vector and axial-vector analogs) \cite{GH}.   

The original NJL model contains the U$_L(2)\times $U$_R(2)$-symmetric and the
U$_A(1)$ breaking terms with equal weights in the Lagrangian, i.e., with $G_1=
G_2.$ This case corresponds to the complete vanishing of interaction in the 
isoscalar pseudoscalar and isovector scalar channels, and hence to the complete
disappearance of these states from the spectrum of the model, as seen in Eq. 
(17). This situation is referred to as ``maximal U$_A(1)$ breaking'' in ref.
\cite{Dmitra}. Since isoscalar pseudoscalar mesons $\eta $ and $\eta ^{'}$ do
exist and one can construct the ideal mixture of them, i.e., a linear
combination of the two that contains no $s\bar{s}$ component (and corresponds
to the isoscalar pseudoscalar state of the two-flavor version of NJL), one has
to relax the severity of U$_A(1)$ breaking in the model and thus consider the
``minimally extended'' \cite{Dmitra} two-flavor NJL model, Eq. (17), with $G_1
\neq G_2.$

Working out this model in a standard manner, one finds the familiar NJL gap
equation for the constituent quark mass $m$ \cite{Klev},
\bqry
m & = & m^0\;-\;\left( G_1+G_2\right) \langle \bar{\psi }(x)\psi (x)\rangle _0
 \NL
 & = & m^0\;+\;4iN_cN_f\left( G_1+G_2\right) \int \frac{d^4p}{(2\pi )^4}\;
\frac{m}{p^2-m^2},
\eqry
which has to be regularized either by introducing a Euclidean cutoff or 
following Pauli and Villars (PV) \cite{PV}. The self-consistency condition 
$\Sigma _H=m,$ where $\Sigma _H$ is the quark self-energy, determines $m$ at 
the one-loop level. This $m$ is also related to the quark condensate:
$$ m=m^0-\left( G_1+G_2\right) \langle \bar{\psi }(x)\psi (x)\rangle _0.$$
A nonzero value of $m$ in the chiral limit $m^0=0$ signals the breakdown of
chiral symmetry.

The meson masses are further read off from the poles of the corresponding 
propagators of the inhomogeneous Bethe-Salpeter equation describing the 
scattering of quarks and antiquarks which are
\beq
-iD(k)=\frac{i(G_1+G_2)}{1-(G_1+G_2)\Pi (k)},
\eeq
where $k$ is the four-momentum transfer, and $\Pi (k)$ represents the sum of 
all proper polarization diagrams in the relevant channel. The form of the 
interaction in Eq. (17) gives rise to scattering in the four channels: the 
isovector pseudoscalar $(\pi ),$ isoscalar pseudoscalar $(\eta _n),$ 
isoscalar scalar $(\sigma )$ and isovector scalar $(\mbox{\boldmath $\sigma 
$}).$ One finds, e.g. \cite{Klev},
\beq
\Pi _\pi (k)-\frac{\Sigma _H}{(G_1+G_2)m}=-4iN_ck^2I(k),
\eeq
where $I(k)$ is a logarithmically divergent loop integral,
\beq
I(k)=\int \frac{d^4p}{(2\pi )^4}\;\frac{1}{[p^2-m^2][(p+k)^2-m^2]},
\eeq
which is PV-regularized. It follows from (20) that
$$\Sigma _H=\left( G_1+G_2\right) m\Pi _\pi (0)$$ which, when combined with
$\Sigma _H=m,$ leads to the self-consistency condition
\beq
1=\left( G_1+G_2\right) \Pi _\pi (0).
\eeq
Then $D$ of Eq. (19) for the pion mode becomes, through (20) \cite{Dmitra}
\bqry
-iD_\pi (k) & = & \frac{i(G_1+G_2)}{1-(G_1+G_2)\Pi _\pi (k)}=\frac{1}{4N_cI(
k)(k^2+i\varepsilon )} \NL
& = & \frac{-ig^2_{\pi qq}}{(k^2+i\varepsilon )F(k)},
\eqry
and similarly, for the remaining three modes \cite{Dmitra}, 
\bqry
-iD_\sigma (k) & = & \frac{i(G_1+G_2)}{1-(G_1+G_2)\Pi _\sigma (k)}\;=\;\frac{
-ig^2_{\pi qq}}{(k^2-4m^2+i\varepsilon )F(k)}, \\
-iD_{\eta _n}(k) & = & \frac{i(G_1-G_2)}{1-(G_1-G_2)\Pi _\pi (k)}\;=\;\frac{
-ig^2_{\pi qq}}{(k^2+i\varepsilon )F(k)-m^2_{tH}}, \\
-iD_{\mbox{\boldmath $\sigma $}}(k) & = & \frac{i(G_1-G_2)}{1-(G_1-G_2)\Pi _
\sigma (k)}\;=\;\frac{-ig^2_\pi qq}{(k^2-4m^2+i\varepsilon )F(k)-m^2_{tH}},
\eqry
where
\beq
m^2_{tH}\left( N_f=2\right) \;\equiv \;\frac{2g^2_{\pi qq}G_2}{G_1^2-G_2^2}\;
\simeq \;\frac{2g^2_{\pi qq}G_2}{G_1^2}\;\!+\;\!O\left( \frac{1}{N_c^2}\right) 
\eeq
is the  't Hooft mass \cite{Dmitra}, and the associated zero external momentum
coupling constants are \cite{Dmitra}
\bqry
g^2_{\pi qq} & = & g^2_{\sigma qq}\;=\;g^2_{\eta _nqq}\;=\;g^2_{\mbox{
\boldmath $\sigma $}qq}\;=\;\left( \frac{\partial \Pi }{\partial k^2}\right) ^{
-1}_0 \NL
& = & \Big[ -4iN_cI(0)\Big] ^{-1}\;=\;\left( \frac{m}{f_\pi }\right) ^2,
\eqry
where $f_\pi =93$ MeV is the pion decay constant. In Eqs. (23)-(27), $F(k)=I(
k)/I(0).$ This factor $F(k)$ provides the $D$'s with a more complicated 
analytic structure than that of the free scalar propagator; this reflects the 
composite nature of the bound states which they describe. However, as 
discussed in detail in ref. \cite{Dmitra}, one may set $F(k)=1$ as a reliable 
approximation which does not affect the predictions of the model essentially. 
With $F(k)=1,$ reading off the poles of the corresponding $D$'s in (23)-(27), 
and switching to the hadron spectroscopy notations $$f_{0n}=\sigma ,\;\;\;a_0=
\mbox{\boldmath $\sigma $},$$ one finds the following meson masses, upon 
introducing explicit chiral symmetry breaking in the form of nonzero current 
quark masses $m^0$ \cite{Dmitra}:
\bqry
m_{\eta _n}^2 & = & m_\pi ^2\;+\;m_{tH}^2\left( N_f=2\right) , \\
m_{a_0}^2 & = & m_\pi ^2\;+\;4m^2\;+\;m_{tH}^2\left( N_f=2\right) , \\
m_{f_{0n}}^2 & = & m_\pi ^2\;+\;4m^2,
\eqry      
which lead to the mass relation
\beq
m_{a_0}^2-m_{f_{0n}}^2=m_{\eta _n}^2-m_\pi ^2=m_{tH}^2\left( N_f=2\right) ,
\eeq
found first by Dmitrasinovic \cite{Dmitra}, and which is a direct consequence
of U$_A(1)$ breaking in this NJL model (by instanton-induced 't Hooft 
interaction). Here $\eta _n$ and $f_{0n}$ are the nonstrange ideal mixtures of
the $\eta ,\eta ^{'}$ and $f_0,f_0^{'}$ mesons, respectively, in the badly 
broken SU(3) limit. 

It can now be easily understood how ``maximal breaking'' of U$_A(1)$ occurs in
this NJL model. It is reached in the limit of equal coupling constants, $G_1=
G_2.$ In this limit, in which the simplest NJL Lagrangian is recovered, as
seen in Eq. (17), the 't~Hooft mass, Eq. (27), and the masses of $\eta _n$ and
$a_0,$ go to infinity. Hence, these heavy modes cannot propagate, so that they
completely decouple from the model. That explains their absence in the original
NJL model. 

The three-flavor generalization of this NJL model is straightforward: The free
Lagrangian and the U$(3)_L\times $U$(3)_R$-symmetric quartic self-interaction
terms are essentially the same as in Eq. (17), the number of terms being 
appropriately extended to 18:
\bqry
L_{NJL}^{(6)} & = & \bar{\psi }\left[ i\gamma \partial -m^0\right] \psi \;+\;
G\sum _{i=0}^8\Big[ \left( \bar{\psi }\mbox{\boldmath $\lambda $}_i\psi \right)
^2+\left( \bar{\psi }i\gamma _5\mbox{\boldmath $\lambda $}_i\psi \right) ^2
\Big]   \NL
 &   & -\;K\Big[ {\rm det}\left( \bar{\psi }(1+\gamma _5)\psi \right) +{\rm 
det}\left( \bar{\psi }(1-\gamma _5)\psi \right) \Big] ,
\eqry 
where $\mbox{\boldmath $\lambda $}_i$ are the SU(3) Gell-Mann structure 
constants, but the U$_A(1)$ breaking determinant interaction term is now of 
sixth order in the Fermi fields, rather than of fourth order as in the $N_f=2$
case. Since one cannot work directly with a sixth-order operator, one has to 
construct an effective mean-field quartic Lagrangian using the expectation 
values of $\bar{\psi }\psi $ \cite{Klev,BJM}. In the $SU(3)$-symmetric limit, 
$\langle \bar{\psi }\mbox{\boldmath $\lambda $}_0\psi \rangle =\sqrt{2/3}\;\!
\langle \bar{\psi }\psi \rangle \neq 0,$ $\langle \bar{\psi }\mbox{\boldmath 
$\lambda $}_3\psi \rangle =\langle \bar{\psi }\mbox{\boldmath $\lambda $}_8
\psi \rangle =0,$ one finds the following effective Lagrangian \cite{Klev}:
\bqry
L_{eff}^{(4)} & = & \bar{\psi }\left[ i\gamma \partial -m^0\right] \psi \;+\;
\Big[ K_0^{(-)}\left( \bar{\psi }\mbox{\boldmath $\lambda $}_0\psi \right) ^2+
\sum _{i=1}^8K_i^{(+)}\left( \bar{\psi }i\gamma _5\mbox{\boldmath $\lambda $}_i
\psi \right) ^2\Big] \NL 
 &   & +\;\Big[ K_0^{(+)}\left( \bar{\psi }i\gamma _5\mbox{\boldmath $\lambda 
$}_0\psi \right) ^2+\sum _{i=1}^8K_i^{(-)}\left( \bar{\psi }\mbox{\boldmath 
$\lambda $}_i\psi \right) ^2\Big] ,
\eqry 
where
\bqry
K_0^{(\pm )} & = & G\;\pm \;K\langle \bar{q}q\rangle \;=\;\frac{1}{2}\left( G_
1\;\mp \;2G_2\right) , \\ 
K_i^{(\pm )} & = & G\;\mp \;\frac{1}{2}K\langle \bar{q}q\rangle \;=\;\frac{1}{
2}\left( G_1\;\pm \;G_2\right) ,\;\;\;i=1,2,\ldots ,8,
\eqry
i.e.,
\beq
G_1=2G,\;\;\;G_2=-\frac{1}{3}K\langle \bar{q}q\rangle ,
\eeq
where the quark condensates are defined as
\bqry
\langle \bar{q}q\rangle  & = & -iN_c\;{\rm Tr}\;\!S_F(x,x)^q\;=\;-4iN_c\int 
\frac{d^4p}{(2\pi )^4}\;\frac{m_q}{p^2-m_q^2+i\varepsilon },\;\;\;q=u,d,s, \NL
\langle \bar{\psi }\psi \rangle & = & \langle \bar{u}u\rangle +\langle \bar{d}
d\rangle +\langle \bar{s}s\rangle \;=\;-iN_c\;{\rm Tr}\;\!S_F(x,x).
\eqry

As in the $N_f=2$ case, the meson masses are read off from the poles of the
corresponding propagators which are in turn constrained by the system of the 
gap equations \cite{Dmitra}:
\bqry
m_u & = & m_u^0\;-\;4G\langle \bar{u}u\rangle \;+\;2K\langle \bar{d}d\rangle
\langle \bar{s}s\rangle , \NL
m_d & = & m_d^0\;-\;4G\langle \bar{d}d\rangle \;+\;2K\langle \bar{s}s\rangle
\langle \bar{u}u\rangle , \NL
m_s & = & m_s^0\;-\;4G\langle \bar{s}s\rangle \;+\;2K\langle \bar{u}u\rangle
\langle \bar{d}d\rangle .
\eqry
In the chiral limit when the bare (current) masses $m_q,\;q=u,d,s$ vanish, and 
in the SU(3)-symmetric limit where $m_u=m_d=m_s,$ one has $\langle \bar{u}u
\rangle =\langle \bar{d}d\rangle =\langle \bar{s}s\rangle =\langle \bar{q}q
\rangle ,$ and hence the system (39) decouples:
\beq
m_q=m_q^0-4G\langle \bar{q}q\rangle +2K\langle \bar{q}q\rangle ^2=m_q^0-4K_1^{
(+)}\langle \bar{q}q\rangle .
\eeq
Upon introducing explicit chiral breaking in the form of nonzero current quark 
masses $m_i^0,$ we find the following results from Eq. (40) and the effective 
Lagrangian (34) \cite{Dmitra}:
\bqry
m_{a_0}^2 & = & m_\pi ^2\;+\;4m^2\;+\;\frac{2}{3}\;\!m_{tH}^2\left( N_f=3
\right) ,  \\
m_{K^\ast _0}^2 & = & m_K^2\;+\;4m^2\;+\;\frac{2}{3}\;\!m_{tH}^2\left( N_f=3
\right) ,  \\
m_{\eta }^2\;+\;m_{\eta ^{'}}^2 & = & 2m_K^2\;+\;m_{tH}^2\left( N_f=3\right) ,
 \\
m_{f_0}^2\;+\;m_{f_0^{'}}^2 & = & 2m_{K_0^\ast }^2\;-\;m_{tH}^2\left( N_f=3
\right) ,  \\
m_{f_0}^2\;+\;m_{f_0^{'}}^2 & = & m_{\eta }^2\;+\;m_{\eta ^{'}}^2\;+\;8m^2\;-\;
\frac{2}{3}\;\!m_{tH}^2\left( N_f=3\right) ,
\eqry      
where
\beq
m^2_{tH}\left( N_f=3\right) \;\equiv \;\frac{3g^2_{\pi qq}G_2}{G_1^2-G_2^2}\;=
\;\frac{3}{2}\;\!m^2_{tH}\left( N_f=2\right) \;\simeq \;\frac{3g^2_{\pi qq}G_
2}{G_1^2}\;\!+\;\!O\left( \frac{1}{N_c^2}\right) , 
\eeq
in view of (27).

Summing up Eqs. (44) and (45), one immediately obtains the Dmitrasinovic sum
rule (4). This sume rule is not, however, the only consequence of this model.
Indeed, Eqs. (32),(43),(46) and $m_{\eta }^2+m_{\eta ^{'}}^2=m_{\eta _n}^2+m_{
\eta _s}^2$ which follows in general from (11), and in particular, with the
mixing angle (10), from (13),(14), lead to
\bqry
m_{\eta _n}^2\;-\;m_\pi ^2 & = & \frac{2}{3}\;\!m_{tH}^2\left( N_f=3\right) ,
 \\
m_{\eta _n}^2\;+\;m_{\eta _s}^2 & = & 2m_K^2\;+\;m^2_{tH}\left( N_f=3\right) ,
\eqry  
which, upon eliminating $m_{tH}^2(N_f=3),$ yields
\beq
2m_{\eta _s}^2+3m_\pi ^2=4m_K^2+m_{\eta _n}^2.
\eeq
Similarly, Eqs. (32),(44),(46) and 
\beq
m_{f_0}^2+m_{f_0^{'}}^2=m_{f_{0n}}^2+m_{f_{0s}}^2,
\eeq
which follows from Eqs. (55),(56) below, lead to
\bqryn
m_{a_0}^2\;-\;m_{f_{0n}}^2 & = & \frac{2}{3}\;\!m_{tH}^2\left( N_f=3\right) ,
 \\
m^2_{f_{0n}}\;+\;m^2_{f_{0s}} & = & 2m_{K_0^\ast }^2\;-\;m^2_{tH}\left( N_f=3
\right) ,
\eqryn  
which again, upon eliminating $m_{tH}^2(N_f=3),$ yields
\beq
2m^2_{f_{0s}}+3m_{a_0}^2=4m_{K_0^\ast }^2+m^2_{f_{0n}}.
\eeq
Eqs. (49) and (51) represent the formula (7) written for the pseudoscalar and
scalar mesons, respectively.

We consider the fact that the formula (7) can be derived in (at least) two 
completely independent ways, viz., from Regge phenomenology, as done in ref. 
\cite{BG}, and in the NJL model, as done in the present paper, as sufficient 
to treat this formula as (almost) model-independent.\footnote{We know of at 
least four other distinct approaches which also yield these same results.} 
As discussed above, in the pseudoscalar meson case, this formula reduces to 
Eq. (15) which is satisfied to an accuracy of better than 1\%; it therefore 
finds its direct experimental confirmation in terms of the pseudoscalar meson 
mass spectrum. Since this formula should describe the physical scalar meson 
mass spectrum as well, we shall apply it to the problem of the correct $q\bar{
q}$ assignment for the scalar meson nonet.

We note first that the masses of the ideally mixed counterparts of the physical
$f_0$ and $f_0^{'}$ mesons are easily obtained from Eqs. 
(44),(49),(51):\footnote{Similar relations for the pseudoscalar mesons follow 
from Eqs. (47),(48) above.}
\bqry
m^2_{f_{0n}} & = & m_{a_0}^2\;-\;\frac{2}{3}\;\!m_{tH}^2\left( N_f=3\right) , 
 \\
m^2_{f_{0s}} & = & 2m_{K_0^\ast }^2\;-\;m_{a_0}^2\;-\;\frac{1}{3}\;\!m_{tH}^2
\left( N_f=3\right) ,
\eqry
where the numerical value of $m^2_{tH}(N_f=3)$ is
\beq
m^2_{tH}\left( N_f=3\right) =m_{\eta }^2+m_{\eta ^{'}}^2-2m_K^2\cong 0.72\;{\rm
GeV}^2.
\eeq

To determine the masses of the physical $f_0$ and $f_0^{'}$ states, one has to
know the $f_0$-$f_0^{'}$ mixing angle. 

In the pseudoscalar meson case, the $\eta $-$\eta ^{'}$ mixing angle is given 
by the mass-matrix diagonalization as \cite{BG}
\beq
\tan 2\theta _{\eta \eta ^{'}}=2\sqrt{2}\left( 1-\frac{3m^2_{tH}(N_f=3)}{2(m_
K^2-m_\pi ^2)}\right) ^{-1}\approx -19^o, 
\eeq
in agreement with (10). In the scalar meson case, one has, 
respectively,\footnote{The formula (56) coincides with Eq. (51) of ref. 
\cite{Dmitra}, and Eq. (41) of ref. \cite{Dmitra2}.} 
\beq
\tan 2\theta _{f_0f_0^{'}}=2\sqrt{2}\left( 1+\frac{3m^2_{tH}(N_f=3)}{2(m_
{K_0^\ast }^2-m_{a_0}^2)}\right) ^{-1},
\eeq
since the sign of the instanton-induced 't Hooft interaction (the sign of $m^2
_{tH})$ for the scalar mesons is opposite to that for the pseudoscalar mesons,
as is clear from the above consideration of the NJL model.

As seen in Eqs. (47),(48),(52),(53),(55),(56), when the 't Hooft interaction 
is switched off, $m^2_{tH}(N_f=3)=0,$ one recovers ideal mixing, $\theta _{
\eta \eta ^{'}}=\theta _{f_0f_0^{'}}=\arctan 1/\sqrt{2},$ and the 
corresponding ideal structures for both multiplets, 
$$m^2_{\eta }=m^2_{\eta _s}=2m_K^2-m_\pi ^2,\;\;\;m^2_{\eta ^{'}}=m^2_{\eta _
n}=m^2_\pi ,$$ $$m^2_{f_0}=m^2_{f_{0s}}=2m_{K_0^\ast }^2-m_{a_0}^2,\;\;\;m^2_{
f_0^{'}}=m^2_{f_{0n}}=m^2_{a_0},$$
which are solutions to Eqs. (47) and (49), respectively, as they should be.

Using Eqs. (52),(53),(56) and
\bqry
m^2_{f_{0n}} & = & \sin ^2\xi \;m^2_{f_0}\;+\;\cos ^2\xi \;m^2_{f_0^{'}}, \\
m^2_{f_{0s}} & = & \cos ^2\xi \;m^2_{f_0}\;+\;\sin ^2\xi \;m^2_{f_0^{'}},
\eqry 
\beq
\cos \xi =\frac{\sin \theta _{f_0f_0^{'}}+\sqrt{2}\cos \theta _{f_0f_0^{'}}}{
\sqrt{3}},
\eeq
which follow from relations similar to Eqs. (11),(12) in the scalar meson 
case, one can easily calculate the masses of the $f_0$ and $f_0^{'}$ states 
and their mixing angle.

The results of our calculation are presented in Table I where we have 
considered a number of different possibilities for the mass of the isovector 
scalar state $a_0$ consistent with the masses of the experimental candidates
$a_0(980),$ $a_0(1320)$ and $a_0(1450).$

\begin{center}
\begin{tabular}{|c|c|c|c|} \hline
$m_{a_0},$ MeV & $m_{f_0},$ MeV & $m_{f_0^{'}},$ MeV & $\theta _{f_0f_0^{'}},$
degrees  \\ \hline \hline
 983.5 & 1710 &  663 &  27.3 \\ \hline \hline
  1290 & 1511 & 1039 &  18.1 \\ \hline
  1320 & 1490 & 1069 &  15.8 \\ \hline
  1350 & 1471 & 1096 &  12.8 \\ \hline \hline
  1410 & 1437 & 1140 &   3.8 \\ \hline
  1450 & 1424 & 1156 &  -4.8 \\ \hline
  1490 & 1423 & 1157 & -14.6 \\ \hline
\end{tabular}
\end{center}
{\bf Table I.} Predictions for the masses and mixing angle of the scalar mesons
for different values of the $a_0$ meson mass, according to Eqs. 
(52),(53),(56)-(59). The value of the $K_0^\ast $ meson mass used in the
calculation is $m_{K_0^\ast }=1429$ MeV. \\

For the $a_0(980)$ taken as the isovector scalar state, our solution exhibits 
a number of interesting features: It predicts an $f_0$-$f_0^{'}$ mixing angle
close to the ideal mixture value (8), i.e., both isoscalar states would be 
almost pure $n\bar{n}$ and $s\bar{s}$; their masses are predicted to be in 
agreement with those of the $\sigma $ of the most recent edition of PDG and 
the $f_J(1710),$ respectively. The existence of the latter state, which in 
this case is almost pure $s\bar{s},$ would be in agreement with the prediction
by Amsler and Close \cite{AC}. However, we will disregard this solution as 
unphysical. Indeed, as we discuss at the beginning of the paper, convincing
arguments exist in the literature regarding the $a_0(980)$ not being a $q\bar{
q}$ state, but rather molecular in character. The existence of the 
$\sigma $-meson (to be identified as the $f_0^{'}(663)$ in this case) is 
controversial due to its enormous width. The spin of the $f_J(1710)$ meson is 
still uncertain $(J=0$ or 2). Moreover, with this solution, the mass interval 
occupied by the scalar nonet turns out to be wider than 1 GeV, a typical mass 
scale of light hadron spectroscopy, which cannot be considered as plausible.

Thus, we restrict ourselves to the solution with $m_{a_0}=1390\pm 100$ MeV (to
accommodate the both experimental candidates $a_0(1320)$ and $a_0(1450))$ 
which, according to the results presented in Table I, is
$$m_{f_0}=1467\pm 44\;{\rm MeV,}\;\;\;m_{f_0^{'}}=1098\pm 59\;{\rm MeV,}\;\;\;
\theta _{f_0f_0^{'}}=1.8\pm 16.3^o.$$

It is interesting to note that for rather wide ranges of the $a_0$ mass and 
$f_0$-$f_0^{'}$ mixing angle, the masses of two isoscalar states are 
accurately predicted to be in the vicinity of 1.1 GeV and 1.45 GeV. Especially
if the $a_0(1450)$ is the true isovector scalar state (which we hope will be
either confirmed or refuted by ongoing experiments), then for a wide range of
its mass $(1450\pm 40$ MeV) the solution picks the masses of the two isoscalars
very accurately at $1430\pm 7$ MeV and $1148\pm 8$ MeV. Nothing can be said,
however, about their decay properties so far, since the solution, albeit 
specifying the two masses, cannot choose between the mixing angles, the proper
determination of which requires a specific value of the $a_0$ mass. 

The solution that we obtained provides the mass of the scalar isoscalar mostly 
singlet state around 1.1 GeV. This value may be considered as the pole position
which gets shifted further down due to strong coupling to, e.g., the $K\bar{
K}$ threshold. For example, recent analysis of various experimental data by 
Anisovich {\it et al.} \cite{Anis} reveals a lowest mass scalar resonance with
the pole position at $1015\pm 15$ MeV and width of $43\pm 8$ MeV. We do not 
address the question of choosing between the $f_0(980)$ and $f_0(1000)$ in this
paper. Let us only mention that with the $f_0$-$f_0^{'}$ mixing angle $\theta
_{f_0f_0^{'}}=-1/2\;\arctan 1/(2\sqrt{2})\approx -9.7^o$ which is achieved with
$m_{a_0}\approx 1470$ MeV, the quark content of the two isoscalars turn out to
be $$f_0=\frac{n\bar{n}-s\bar{s}}{\sqrt{2}},\;\;\;f_0^{'}=\frac{n\bar{n}+s\bar{
s}}{\sqrt{2}},\;\;\;n\bar{n}\equiv \frac{u\bar{u}+d\bar{d}}{\sqrt{2}},$$ i.e.,
the both have (50\% $n\bar{n},$ 50\% $s\bar{s})$ quark content. Such quark
content can, e.g., easily explain a large branching ratio of 22\% of the $f_0(
980)$ decay into the $K\bar{K}$ channel. 

The mass of the scalar isoscalar mostly octet state predicted by our solution,
$1467\pm 44$ MeV, is in good agreement with recent analysis by Kaminski {\it 
et al.} \cite{KLL} of the three coupled channel $(\pi \pi,\;K\bar{K},\;4\pi )$
model which reveals a relatively narrow $(\Gamma =135\pm 45$ MeV) resonance 
with a mass of $1430\pm 30$ MeV.   

We remark also that the masses of both isoscalar scalar states predicted in 
this paper are below a typical range of $1600\pm 100$ MeV provided by QCD 
lattice calculations as that for the lightest scalar glueball 
\cite{SVW,Luo,lat}. They may however get shifted from the predicted values 
because of possible mixture with the lightest scalar glueball. 
        
The $q\bar{q}$ assignment found in this work,
\beq
a_0(1320)\;\;{\rm or}\;\;a_0(1450),\;\;\;K_0^\ast (1430),\;\;\;f_0(1450),\;\;\;
f_0(980)\;\;{\rm or}\;\;f_0(1000),
\eeq
is consistent with the assignments (3),(5),(6) established in three different
approaches. The fact that different approaches lead to essentially the same 
result encourage us to believe that we are not far from the complete resolution
of the scalar meson nonet enigma.

\end{document}